\documentclass[prd,twocolumn,noeprint]{revtex4-1}
\usepackage{graphicx,natbib}
\usepackage{url}
\usepackage{ulem}
\usepackage{color}
\usepackage{rotating}
\usepackage{amssymb,amsmath}

\newcommand{\fnl}{f_{\rm NL}}
\newcommand{\fnlmink}{f_{\rm NL}^{\rm MF}}
\newcommand{\fnlbisp}{f_{\rm NL}^{\rm bisp}}
\newcommand{\fnlbar}{\bar{f}_{\rm NL}}

\newcommand{\fsky}{f_{\rm sky}}

\newcommand{\mnras}{{Mon.~Not.~R.~Astron.~Soc.}}
\newcommand{\apjs}{{Astrophys.~J.~Supp.}}

\newcommand{\jcap}{{J. Cosmol. Astropart. Phys.}}

\newcommand{\wj}[6]{\left(
                           \begin{array}{ccc}
        \! #1\! & #2\!  & #3\!  \\
        \! #4\! & #5\!  & #6\!
                           \end{array}
                   \right)}

\newcommand{\nhat}{\mathbf{\hat{n}}}

\begin{document}

\title{Joint Minkowski Functionals and Bispectrum Constraints on
Non-Gaussianity in the CMB}

\author{Wenjuan Fang}
\email{wjfang@illinois.edu}
\affiliation{Department of Astronomy, University of Illinois at Urbana-Champaign, 1002 W. Green St, Urbana, IL 61801}
\affiliation{Department of Physics, University of Michigan, 
450 Church St, Ann Arbor, MI 48109-1040}

\author{Adam Becker}
\email{beckeram@umich.edu}
\affiliation{Department of Physics, University of Michigan, 
450 Church St, Ann Arbor, MI 48109-1040}

\author{Dragan Huterer}
\email{huterer@umich.edu}
\affiliation{Department of Physics, University of Michigan, 
450 Church St, Ann Arbor, MI 48109-1040}

\author{Eugene A. Lim}
\email{eugene.a.lim@gmail.com}
\affiliation{Theoretical Particle Physics and Cosmology Group, Physics Department,
Kings College London, Strand, London WC2R 2LS, United Kingdom}
\affiliation{Centre for Theoretical Cosmology, Department of Applied Mathematics and Theoretical Physics,
University of Cambridge, Wilberforce Road, Cambridge CB3 0WA, United Kingdom}

\begin{abstract} 
Two of the most commonly used tools to constrain the primordial
non-Gaussianity are the bispectrum and the Minkowski functionals of CMB
temperature anisotropies. These two measures of non-Gaussianity in principle
provide distinct (though correlated) information, but in the past constraints
from them have only been loosely compared and not statistically combined. In
this work we evaluate, for the first time, the covariance matrix between the
local non-Gaussianity coefficient $\fnl$ estimated through the bispectrum and
Minkowski functionals. We find that the estimators are positively correlated,
with correlation coefficient $r\simeq 0.3$. Using the WMAP7 data to combine
the two measures and accounting for the point-source systematics, we find the
combined constraint $\fnl=37\pm 28$, which has a $\sim 20\%$
smaller error than either of the individual constraints.
\end{abstract}

\date{\today}

\maketitle

\section{Introduction}  

Detection of any departures from Gaussianity in the
distribution of primordial fluctuations would give important information about
inflation.  Primordial non-Gaussianity (henceforth NG) imprints signatures on the cosmic microwave
background (CMB) and large-scale structure, and these cosmological probes
can in turn provide excellent constraints on primordial NG and thus inflationary models; for
reviews, see \cite{Komatsu_CQG,Bartolo:2004if, Liguori_AA,Yadav_Wandelt_AA}.

Two of the principal statistics on the CMB used to constrain NG are the
bispectrum (harmonic transform of the three-point correlation function) of the
CMB temperature fluctuations, and Minkowski functionals (henceforth MF) which
roughly measure the connectedness or morphology of the CMB field. In the
``local'' model of NG, the primordial curvature perturbation $\Phi(\bf{x})$
has a quadratic term correction: $\Phi = \Phi_G + \fnl(\Phi_G^2 -
\langle\Phi_G^2\rangle )$, where $\Phi_G$ is an auxiliary Gaussian field
\cite{Komatsu_Spergel}.  Recent constraints obtained on the non-linear
coupling constant $\fnl$ using the WMAP data are $\fnl=37\pm 20$ from the
bispectrum analysis \cite{WMAP9} (see also
\cite{Yadav_Wandelt_evidence,Smith_Sen_Zal_optimal}), and $\fnl=20\pm42$ from
the MF analysis~\cite{HikMat12}.

Since MF are morphological statistics, they probe NG both in configuration
space and to all orders of the statistics of the temperature anisotropy
field. This means that they sample the anisotropy map differently from the
usual bispectrum (and higher order polyspectra) measurements, albeit in a
suboptimal way -- this fact is crucial as joint constraints will in principle
yield different constraints. Furthermore, unlike the bispectrum estimators
which require a template (i.e.\ $k$-space configuration with a free amplitude)
such as the local, equilateral or orthogonal type, MF are in principle
template-free, although in practice one can construct a template-based MF
estimator as we have done in this paper.

In principle, the MF are sensitive to the weighted sum of the bispectrum
coefficients (out to the smallest scale measured) \cite{HikKomMat06}, so the
MF would naively be expected to contain only a subset of the same information
as the bispectrum. In reality, however, this idealized expectation is not
borne out: the bispectrum and the MF partially complement each other, and
their information is not 100\% correlated. One reason for this is the fact
that the optimal bispectrum estimators
\cite{Babich:2005en,Creminelli_limits_WMAP} are
computationally challenging to implement for current high-precision CMB
experiments \cite{Smith_Sen_Zal_optimal,Planck_NG}, and they are anyway only
optimal for the case of vanishing non-Gaussianity
\cite{Creminelli_estimators,Liguori:2007sj}. Moreover, the bispectrum and MF
are sensitive to different astrophysical and analysis-related systematics,
given that they are defined in the harmonic and real space
respectively. Hence, combining the constraints obtained by current fast though sub-optimal 
bispectrum estimators with those from the MF, as we do in this paper, provides
an alternative to improving the ``optimality'' of these estimators, and makes
the combined constraints both stronger and more robust.

Hence an obvious question is {\it how correlated} are the MF and the
bispectrum estimators, and consequently what is the combined constraint on NG from
them. This is the question that we address in this paper -- we will show that
the correlation between the two estimators, while nonzero, is far from
maximal.  Having calculated that, we compute the joint estimate of NG from
both statistics.

\medskip
\section{Minkowski functionals methodology}  

The three Minkowski functionals
$V_i\ (i=0,1,2)$ describe morphological properties of the hot and cold spots
in the CMB temperature map. The morphology of the map, and thus the MF, are
studied by specifying a temperature threshold $\nu\equiv(\Delta T/T)/\sigma_0$
in the map, where $\sigma_0$ is the rms of the fractional temperature
fluctuation $\Delta T/T$, hereafter simply denoted as $f$. Specifically, $V_0$
is the area fraction of the regions above the temperature threshold, $V_1$ is
their boundary length, and $V_2$ is the geodesic curvature integrated along
their boundary, which in a compact $S_2$ space is related to the Euler
characteristic $\chi$ by $\chi=V_2+V_0/2\pi$~\cite{SchGor98}.  The MF can be
expressed as integrals of functions of the anisotropy field and its
derivatives over the compact space of the CMB sky. For explicit expressions
see e.g.\ \cite{SchGor98,HikKomMat06}; we shall adopt these operationally
convenient forms to calculate the MF for a given map.

If the temperature fluctuations are Gaussian, the ensemble
averages of the Minkowski functionals have analytic expressions that are
completely specified by the two-point statistics (variance) of the
fluctuations,  $\sigma_0^2$ and $\sigma_1^2(\equiv\langle|\nabla
f|^2\rangle)$~\cite{Tomita86}. On the other hand, when the fluctuations are
weakly non-Gaussian and the cumulants $\langle f^n\rangle_c$ (where ``c''
stands for the connected part) satisfy the hierarchical ordering $\langle
f^n\rangle_c \sim \sigma_0^{2n-2}$, one can obtain an order-by-order expansion
in powers of $\sigma_0$ for the average of the Minkowski
functionals~\cite{Mat03,Mat10}. In this letter, we consider the first order in
the hierarchical non-Gaussian expansion, which in addition to $\sigma_0$ and
$\sigma_1$ depends on the three-point statistics (skewness) of the field:
$S\equiv\langle f^3\rangle,\ S_I\equiv\langle f^2\nabla^2f
\rangle,\ S_{II}\equiv2\langle |\nabla f|^2\nabla^2 f \rangle$. The two
variance and the three skewness parameters can be calculated from theory by integrating
over the power spectrum and bispectrum of the CMB field, respectively; for
explicit expressions, see~\cite{HikKomMat06,HikMat12}. In the special case of
the local-type primordial non-Gaussianity, $S,\,S_{I},\,S_{II}$ are all linearly
proportional to $\fnl$.


In this Letter, we use the co-added V+W band data from the WMAP seven-year
results \cite{wmap7} to obtain our constraints on $\fnl$. The V
and W bands are chosen for they are the most foreground-free. For this purpose, we
generate 1000 simulations of the WMAP data following the procedure given in
Appendix A of~\cite{wmap5}. The only difference (aside from using the WMAP7
cosmological model) is that we used a uniform weighting for the maps,
rather than the slightly more complicated weighting given there, since it only
gives a marginal improvement in estimating $\fnl$. Each of our simulated map is the sum of three
components: 1) the Gaussian CMB
realizations (the ``signal'')  based on the CMB power spectrum calculated assuming
the best-fit WMAP seven-year cosmology including the effect of beam smearing,
2) instrumental noise modeled as the Poisson process with the 
rms noise per pixel $\sigma/\sqrt{N_{\rm obs}}$, where $\sigma$ is the rms
noise per observation and $N_{\rm obs}$ is the number of observations per
pixel, and 3) unresolved point sources modeled as the Poisson realizations from
assuming a single population of sources with a fixed frequency-independent
flux whose flux strength and number density roughly reproduce the source power
spectrum and bispectrum measured from the WMAP Q band. The latter two components
are modeled to closely match the systematics expected in the V+W co-added map. We then mask
both the WMAP data and our simulated maps by using the KQ75 mask.

To make predictions for the ensemble average of the Minkowski functionals when
various observational effects are present, we should also include these
effects in the calculations of the two variance parameters and three skewness
parameters. Each of these parameters has contributions from the noise part --
instrumental noise and point sources, in addition to the beam-smeared CMB
signal part. The noise and signal contributions add up directly since the CMB
signal and noise are uncorrelated. We estimate these noise contributions from
our simulations: we calculate the variance and skewness parameters for each
simulated map, and take their average over the 1000 samples; we then subtract
off the signal contributions which are known to us for these Gaussian CMB
simulations.

Before we proceed to the fitting procedure and obtain our Minkowski functional
constraints on $\fnl$, we address the ``residual problem'' in our numerical
evaluation of the Minkowski functionals. Previous work \cite{Hik08} found
that, even for a set of Gaussian CMB simulations without noise, the averages
of the MF calculated for each map are different from their
values expected from theory. As shown in \cite{LimSim12}, these residuals are
generated by the discrete binning of the MF in the threshold $\nu$, and for weakly non-Gaussian maps can be calculated analytically and then subtracted
order by order in $\sigma_0$. In this work, we instead follow \cite{Hik08}
and calculate the residuals from our simulations as the difference of the
sample-averaged means of the MF and their theoretically expected means. These residuals are then subtracted from the measured Minkowski
functionals. We use the same residuals to account for those for the
non-Gaussian case: for a weakly non-Gaussian field, the differences are at the order of $\sigma_0$.

Before we calculate the Minkowski functionals for each map, we smooth the map
at several different angular scales. This allows us to extract additional
information from the map and tighten the constraints on $\fnl$. Specifically,
we use a Gaussian window function, and smooth each map at five
different scales with the Full Width Half Maximum (FWHM) $\theta$ set at
$\theta=10',\ 20',\ 40',\ 80',\ 100'$. Pixels within
  a distance of $\theta$ away from the boundary of the KQ75 mask are removed
to avoid contamination from the masked regions that may be introduced due to
the smoothing. Ideally, one may want to smooth the maps at
  infinitely many scales and extract the constraint on $\fnl$ by integrating
  over them. Clearly, this cannot be done in reality. The
  five smoothing scales we choose range from roughly the resolution of the
  WMAP V+W band data to the scale at which only $\sim 40\%$ of the map remains
  for analysis. (Note, the larger the smoothing scale, the bigger the area to
  be removed to avoid contamination.) Combining the results at the five
  smoothing scales allows us to recover most of the available information.
For each smoothed map, we calculate its three Minkowski functionals at 15
temperature thresholds from $\nu=-3.5$ to $3.5$ with equal bin size of
$\Delta\nu=0.5$.

To obtain the constraints on $\fnl$, we perform a $\chi^2$ analysis, which
compares theoretical predictions at a given $\fnl$ to the measurements, and is
calculated as
\begin{equation}
\chi^2=\sum_{i,j} \left[V_i^{\rm obs}-V_i^{\rm th}(\fnl)\right] 
C^{-1}_{ij} \left[V_{j}^{\rm obs}-V_{j}^{\rm th}(\fnl)\right],
\end{equation}
where $i$ and $j$ run over all combinations of the 15 thresholds, three orders, and five
smoothing scales for the measured Minkowski functionals. Here $V_i^{\rm obs}$
are the ``observed'' numerically evaluated Minkowski functionals, $V_i^{\rm
  th}$ are the theoretically expected averages for the MF (which are functions
of $\fnl$), and $C$ is the covariance matrix for $(V_i,V_j)$ which we
calculate from our simulations as
\begin{equation}
C_{ij}=\langle 
\left (V_i-\langle V_i\rangle\right ) 
\left (V_j-\langle V_j\rangle\right ) 
\rangle_{\rm sim},
\end{equation}
where the angular brackets denote averaging over the 1000 simulated maps. We
then obtain our best-fit value of $\fnl$, henceforth $\fnlmink$, by minimizing the $\chi^2$.

To check that our estimator for $\fnlmink$ is unbiased, we first apply it to
the 1000 simulated maps either for the MF measurements at each smoothing scale
or their combined results. We find that the average of the best-fit values
accurately reproduces the theoretical input in our simulation, i.e.,
$\fnl=0$. Next, we test our estimator on publicly available {\it non-Gaussian}
CMB maps generated with the local-type NG \cite{ElsWan09}, and we again find
negligible bias (1\% or less of the true $\fnl$) in our estimator.

\begin{table}[!t]
\begin{center}
\begin{tabular}{||c | c | ccc||}
\hline\hline
\rule[-2mm]{0mm}{6mm} & \quad $\theta (')$ && $\fnl$ &\\
\hline\hline
 &10 & $\, 71\!\!\!\!\!$&$\pm$&$\!\!\!\!\! 96 \,$ \\
Minkowski &20& $-21\!\!\!\!\!$&$\pm$&$\!\!\!\!\! 52$ \\
Functionals &40& $-2\!\!\!\!\!$&$\pm$&$\!\!\!\!\! 49$ \\
 &80 & $\, 40\!\!\!\!\!$&$\pm$&$\!\!\!\!\! 73 \,$ \\
 &100 & $\, -16\!\!\!\!\!$&$\pm$&$\!\!\!\!\! 92 \,$ \\
\rule[-2mm]{0mm}{5mm}  &all&$29\!\!\!\!$&$\pm$&$\!\!\!\!\! 33$ \\
\hline
\rule[-2mm]{0mm}{6mm} bispectrum &  & $46\!\!\!\!\!\!$&$\pm$&$\!\!\!\!\! 35$\\
\hline
\rule[-2mm]{0mm}{6mm} {\bf{MF + bisp}}  & & $\bf{37}\!\!\!\!\!\!$ & $\bf{\pm}$ &
$\!\!\!\!\! {\bf{28}}$\\
\hline\hline
\end{tabular}
\end{center}
\caption{Constraints on $\fnl$ from the CMB Minkowski functionals, bispectrum,
  and their combination. The analyses use the WMAP 7-year V+W co-added
  map. $\theta$ is the FWHM of the Gaussian beam used to smooth
  the map for the Minkowski functional analysis. }
\label{tab:results}
\end{table}

Finally, we apply our estimator on the co-added V+W band data from the
WMAP. In Table~\ref{tab:results}, we show the constraints on $\fnlmink$ from
smoothing the map at each of the five angular scales, and the
joint constraint from all scales combined, which we quote as our final MF
constraint: $\fnlmink=29\pm 33$. This constraint is consistent
  with that found by Hikage \& Matsubara \cite{HikMat12}, although we improve
  upon their analysis in a couple of ways: 1) we remove the residuals in the
  numerically evaluated MFs using the method from \cite{Hik08}, as opposed to the
  residual removal  based on the work in \cite{LimSim12} which, we
  found,  causes biases in the estimated $\fnlmink$ by $\sim10$. and 2) we
  carefully include point sources in our simulated WMAP maps.

\medskip
\section{Bispectrum methodology} 
With the MF estimator of $\fnlmink$ obtained, we
next develop $\fnlbisp$ -- the estimator from bispectrum. The observed CMB
bispectrum is given by
\begin{equation}
 B_{\ell_1 \ell_2 \ell_3}=\sum_{m_1 m_2 m_3}\wj{\ell_1}{\ell_2}{\ell_3}{m_1}{m_2}{m_3} a_{\ell_1 m_1}a_{\ell_2 m_2}a_{\ell_3 m_3},
\end{equation}
where the matrix is the Wigner-3j symbol, and $a_{\ell m}$ is the spherical
harmonic transform of the temperature anisotropy map. In the local-type NG
model, $B_{\ell_1 \ell_2 \ell_3}$ is linearly proportional to $\fnl$.

We follow the prescription that uses the KSW \cite{KSW} estimator to calculate
$\fnlbisp$ from CMB maps (see also \cite{nfnl} for the exact implementation
that we use). In brief, the KSW is a cubic (in the
temperature field) estimator of non-Gaussianity; it is nearly minimum-variance
and computationally fast, and can straightforwardly deal with partial sky
coverage and inhomogeneous noise. The first ingredient in using KSW is to calculate the Fisher
matrix $F$ corresponding to $\fnl$; for this we need the theoretical
bispectrum $B^{\rm theory}_{\ell_1 \ell_2 \ell_3}$ which can be calculated
with the help of transfer functions from CAMB \cite{CAMB}. Furthermore, KSW
requires filtered maps $A(\nhat, r)$ and $B(\nhat, r)$ from which the skewness
$S$ of the field can be calculated; these filtered maps can be computed using
HEALPix (by way of HealPy) to perform the forwards and backwards spherical
harmonic transforms that are necessary in their computation. Given the
skewness and the Fisher matrix, the KSW estimator for $\fnl$ is
\begin{equation}
\fnlbisp = {S\over F}.
\label{eq:KSW}
\end{equation}
%
To account for the masking of the CMB sky, we make the substitution $S
\rightarrow S_{\rm cut} = S/\fsky + S_{\rm linear}$ \cite{Yadav2008}. $S_{\rm
  linear}$ is an addition to skewness 
and is calibrated to account for partial-sky observations
\begin{align}
S_{\rm linear}  = & - {1 \over \fsky} \int r^2 dr \int d^2 \nhat  
\left[ A(\nhat, r) \langle B_{\rm sim}^2 (\nhat, r) \rangle_{\rm MC} \right. \nonumber
 \\
& \left. + 2B(\nhat, r) \langle A_{\rm sim}(\nhat, r) B_{\rm sim}(\nhat, r) \rangle_{\rm MC} \right].
\end{align}
The subscripted filtered maps $A_{\rm sim}$ and $B_{\rm sim}$ are created from
Python-produced Gaussian Monte Carlo realizations of the cut CMB sky; the
brackets $\langle \rangle_{\rm MC}$ indicate an average over 300 of the maps. The
simulated maps were produced as outlined earlier when we discussed the MF.

Applying the bispectrum/KSW estimator to the co-added V+W band data of the
WMAP, we obtain the constraint on the local NG to be
$\fnlbisp=46\pm35$. The error we obtained is larger than that
  from Ref.~\cite{wmap7} using the same data because we have used a bispectrum
  estimator that is less optimal but much more convenient to evaluate.

\medskip
\section{Combined analysis} 
In addition to obtaining the constraints on $\fnl$
separately from the MF and bispectrum analyses, we would like to combine them
to extract a more stringent and robust result. To make the problem tractable,
we opt to consistently combine the {\it estimators} of $\fnl$ from these two
analyses, rather than attempting to find the covariance between the {\it
  observables}, i.e. the MF and bispectrum themselves. It is a reasonably good
assumption that the two estimators of $\fnl$ satisfy a bivariate Gaussian
distribution, especially near the peak of the distribution (see
Figure~\ref{fig:CovSmooth} below). Let us organize the two estimators into a
row-vector ${\bf \fnl}\equiv [\fnlmink,\,\fnlbisp]$, and let ${\bf C}$ be the
$2\times 2$ covariance matrix for them. Assuming the underlying true value of
$\fnl$ is $\fnlbar$, we can write down the following joint-distribution for
the two $\fnl$ estimators
\begin{equation}
\mathcal{L}\propto |{\bf C}|^{-1/2} 
\exp\left [
-{1\over 2}({\bf \fnl-\fnlbar}) {\bf C}^{-1}
({\bf \fnl-\fnlbar})^T \right ],
\end{equation}
where ${\bf \fnlbar}=[\fnlbar,\,\fnlbar]\equiv\fnlbar\,\mathbb{I}$. Given a
measurement of ${\bf \fnl}$, a best estimate for $\fnlbar$ can be obtained by
maximizing $\mathcal{L}$. Assuming that the covariance matrix does not depend
on $\fnlbar$, we find the following expressions for the best estimate and
variance of $\fnlbar$ from the combined analysis
\begin{equation}
\fnlbar=\frac
{\mathbb{I} \,{\bf C}^{-1}\,{\bf \fnl}^T }
{\mathbb{I} \,{\bf C}^{-1}\,\mathbb{I}^T},
\quad
\sigma^2_{\fnlbar}=\frac
{1}{\mathbb{I} \,{\bf C}^{-1}\,\mathbb{I}^T}.
\label{eq:fnlbar}
\end{equation}

At the same time, by evaluating both $\fnlmink$ and $\fnlbisp$ for the 1000
simulated WMAP maps, we numerically obtain their joint distribution, as shown
in Figure~\ref{fig:CovSmooth}. From this distribution, we can deduce their
correlation. We find that the two estimators of $\fnl$ are positively
correlated, with a correlation coefficient of
$r=0.32\pm0.03$. We are using the MF constraints
  from combining the five smoothing scales, as these are the final interesting
  MF constraints. However, we also find positive correlations between
  $\fnlbisp$ and the MF constraints obtained at each individual smoothing
  scale: specifically, $r$ varies from $0.46$ to $0.2$ when $\theta$ increases
  from $10'$ to $100'$.  The covariances or off-diagonal elements of
${\bf{C}}$ are then $C_{12}=C_{21}=r\sqrt{C_{11}C_{22}}$; recall that we
already found the variances to be $C_{11}=33^2$ and
$C_{22}=35^2$. We find the bivariate Gaussian distribution with
  the derived covariance matrix ${\bf{C}}$ gives a good description of the
  joint distribution of $(\fnlmink,\fnlbisp)$ for the simulated maps: the
  $68\%,\ 95\%$ contours enclose roughly the same percentages ($\pm1\%$) as in
  the simulated maps, and the orientation of the two distributions agree, see
  Figure~\ref{fig:CovSmooth}.

\begin{figure} [t]
\begin{center}
{\includegraphics[angle=0,scale=0.65]{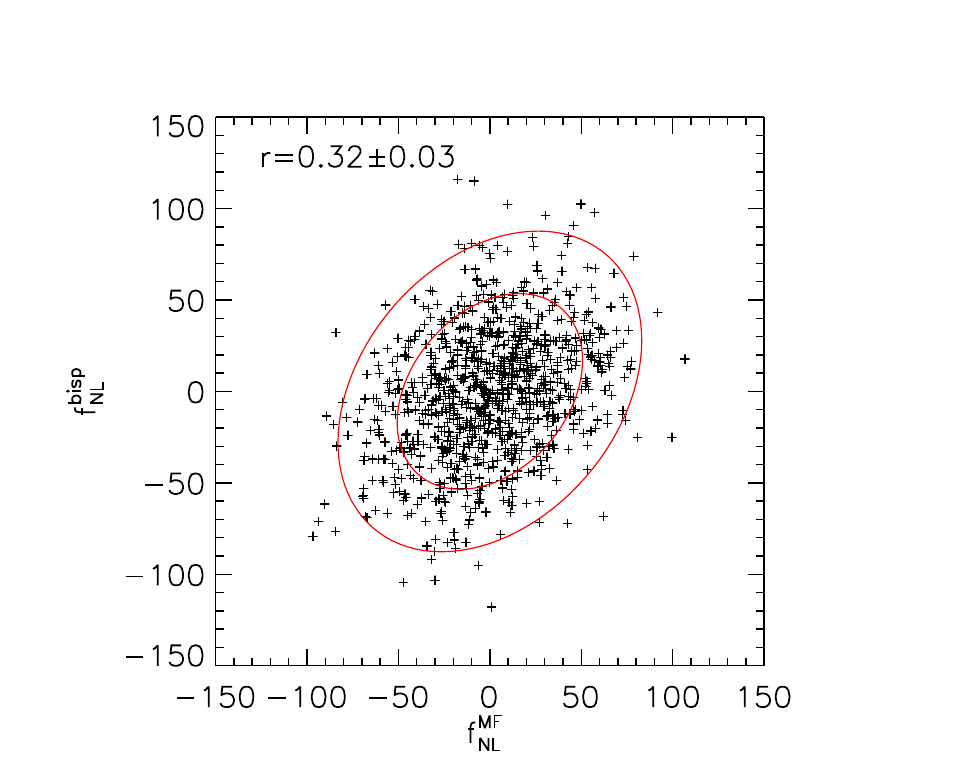}}
 \caption{\label{fig:CovSmooth} Joint-distribution of $\fnlmink$ and
   $\fnlbisp$ from 1000 simulations of the WMAP data including point sources
   and instrumental noise. We find a correlation coefficient of
   $r=0.32\pm0.03$. The $\fnlmink$ estimates are obtained from
   the combination of smoothing the maps at $\theta(\rm
     FWHM)=10',20',40',80',100'$. The contours show the 68\% and 95\%
     confidence regions of the bivariate Gaussian distribution for
     $(\fnlmink,\fnlbisp)$ with its covariance matrix derived from the
     simulations.}
\end{center}
\end{figure}

Using the numerically derived covariance matrix ${\bf C}$, together with
our best-fits for $\fnlmink$ and $\fnlbisp$, we find through Eq.~(\ref{eq:fnlbar}) the combined
constraint to be
\begin{equation}
\fnlbar\equiv \fnl^{\rm MF+bisp}=37\pm28, \label{eq:fnlcomb}
\end{equation}
which has a $\sim 20\%$ improvement in the error with respect to the individual constraints.

\medskip
\section{Conclusions} 
We evaluated, for the first time, the full covariance
matrix for the Minkowski functional estimator of the local-type primordial
non-Gaussianity $\fnlmink$ and the bispectrum estimator $\fnlbisp$. We found
the correlation coefficient $r=0.32\pm 0.03$, and used it to
combine the constraints from the MF and bispectrum (and their respective
variances) to obtain the constraint in Eq.~(\ref{eq:fnlcomb}). 
  Combining these two estimators hence provides an alternative to improving
  their ``optimality'' and leads to combined constraints that are both
  stronger and more robust. Our work can be extended by using more optimal
  estimators, e.g. the bispectrum estimator described in
  \cite{Creminelli_limits_WMAP} whose calculation is numerically very
  challenging, and by applying to the Planck data, which we leave for future
  work.

One convenient feature of
this work is that, by combining the constraints at the level of MF and
bispectrum {\it estimators}, we make the problem tractable: an obvious first
approach could be to calculate the covariance between the observed bispectrum
and MF themselves, but this is extremely complicated, given that the MF and
bispectrum are functions of many scales and/or thresholds. Combining the
different estimators numerically, as we have done here for the case of local
NG, can in principle be rather straightforwardly extended to other types of NG
and other cosmological probes. This type of approach is therefore likely to
become more widespread with new and better data. 

\medskip
\section*{Acknowledgements}  We thank Licia Verde for useful comments. We
acknowledge the use of the publicly available CAMB \cite{CAMB} and HEALPix
\cite{healpix} packages. WF is supported by NASA grant NNX12AC99G. WF, AB and
DH have been supported by the DOE and NSF at the University of Michigan. WF
and DH thank the Aspen Center for Physics, which is supported by the NSF Grant
No.\ 1066293, for the hospitality in the summer of 2012.

\bibliography{fnl}

\end{document}